\documentstyle[preprint,tighten,eqsecnum,prb,aps]{revtex}
\begin{document}
\draft
\title{
Integrable versus Non-Integrable Spin Chain Impurity Models}
\author{
Erik S. S\o rensen$^{(a)}$,
Sebastian Eggert$^{(a)}$, and
Ian Affleck$^{(a,b)}$}
\address{$^{(a)}$Department of Physics and $^{(b)}$Canadian Institute
for Advanced Research}
\address{
University of British Columbia, Vancouver, BC, V6T 1Z1, Canada}
\date{\today\, \, \small UBCTP-93-007}
\maketitle
%\rightline{ \small UBCTP-93-007}
\begin{abstract}
Recent renormalization group studies of impurities in spin-1/2 chains
appear to be inconsistent with Bethe ansatz results for a special
integrable model. We study this system in more detail around the
integrable point in parameter space and argue that this integrable impurity
model corresponds to a non-generic multi-critical point. Using  previous
results on impurities in half-integer spin chains, a consistent
renormalization group flow and  phase diagram is proposed.
\end{abstract}
\section{Introduction}

Recently there has been considerable interest in various quantum impurity
problems.\cite{nuclb,AL1,LA,AL2,KF}  These can generally be
formulated as one-dimensional  Luttinger liquids interacting with  local
defects of various kinds. In general it is expected that such quantum
impurity models renormalize to critical points which correspond to
conformally invariant boundary conditions.  The quantum impurity is
screened and/or decouples; it does not appear in the fixed point
Hamiltonian although a remnant  effective impurity, decoupled from the
continuum degrees of freedom, may be left behind.

 A particularly simple example is a single impurity in a spin  $S=1/2$
Heisenberg antiferromagnetic chain.   Two of the present authors analyzed
a large class of models of this type using analytic renormalization group
(RG) arguments and numerical finite-size analysis~\cite{imp1}. We
concluded that the only stable critical points correspond to a completely
unperturbed chain or else a chain with a break at the impurity location.
Taking the initial boundary conditions to be periodic,
we refer to these two critical points as the periodic and open chain,
respectively.
In the open case, but not the periodic, a remnant impurity spin may
also be present, as we shall review in section~\ref{sec:review}.
It was recently drawn to our attention that a Bethe ansatz
integrable impurity model of this type was solved several years ago.  Its
low-energy behaviour corresponds to a conformally invariant boundary
condition, but not to one of the stable critical points mentioned above.
The purpose of the present article is to resolve this apparent
contradiction.

The integrable impurity
model involves a single spin-$S$ impurity which is coupled
symmetrically to two neighbouring sites on the chain.   The Hamiltonian,
found by Andrei and Johannesson~\cite{andrei} is
\begin{eqnarray}
H_{\rm Int}
&=&\frac{1}{4}\sum_{i=1}^{L-1}{\vec
\sigma}_i\cdot{\vec \sigma}_{i+1} + {1 \over 4} {\vec \sigma}_1 \cdot
{\vec \sigma}_L \nonumber \\
&+&\frac{1}{2}(\frac{2}{2S+1})^2[
{\vec\sigma}_1\cdot {\vec S} +{\vec\sigma}_{L}\cdot {\vec S}
+\frac{1}{2}\{{\vec\sigma}_1\cdot {\vec S},{\vec\sigma}_{L}\cdot {\vec
S}\} -S(S+1){\vec\sigma}_1\cdot{\vec\sigma}_{L}] .
\label{eq:hamil}
\end{eqnarray}
where the $\vec \sigma_i$'s are Pauli matrices and $\{,\}$ denotes
the anti commutator. In the following we shall refer to
Eq.~(\ref{eq:hamil}) as the integrable impurity model. The
equivalent problem of a spin-$1$ chain coupled to a spin-$S$ impurity was
solved by Lee and Schlottmann~\cite{lee}, and later generalized to a
spin-$S'$ chain coupled to a spin-$S$ impurity~\cite{schlottmann}. For $S
\geq S'$ it was found that in the thermodynamic limit the system behaves
like a  spin-$S'$ chain with one extra site and a decoupled spin ${\vec
S}_{\rm eff}$ of size $S-S'$.
The  $S'\geq 1$ models already exhibit
non-generic behaviour before adding the impurity. In particular
these integrable periodic models~\cite{takhtajan,babujian}, without
impurities,
do not exhibit the Haldane gap for integer $S'$.  For a discussion
see Ref.~\onlinecite{AH}. We shall not consider them further.

A peculiar feature of the integrable impurity
model with $S'=1/2$ and $S\geq 1$ is
that the effective, partially screened impurity at the critical point has
spin $S_{\rm eff}=S-1/2$~\cite{schlottmann},
despite the fact that the impurity couples with
equal strength to {\it two} spin-$1/2$'s.
This seems contradictory since,
if we assume that the critical
point corresponds to an infinite antiferromagnetic coupling then we
would obtain $S_{\rm eff}=S-1$.  Furthermore,
our RG analysis indicates that in
any event if $S_{\rm eff}\neq 0$ a stable critical point must correspond to
the {\it open} chain.  Our further analysis of the integrable impurity
model with
$S=1$, discussed in Section III, indicates that the critical point
corresponds to the {\it periodic} chain with  $S_{\rm eff}=S-1/2=1/2$.
It is as if the impurity spin ``splits in half'', donating an extra
$S=1/2$ spin to the periodic chain and leaving behind a decoupled $s=1/2$
impurity. [See Fig.~\ref{fig:split}.]

We argue below that this corresponds to an unstable critical point which
is peculiar to this Hamiltonian.  Generic Hamiltonians renormalize to the
stable fixed points mentioned above.  Thus it appears that the conditions
for integrability somehow ``fine-tune'' the impurity-model Hamiltonian so
that it corresponds to an unstable fixed point.  The same phenomenon was
found earlier for integrable periodic chains of spin $S'\geq 1$~\cite{AH}.

In the next section we briefly review our RG analysis which shows that the
integrable impurity
model cannot correspond to any of the known stable fixed
points. We then conjecture that it corresponds to the particular unstable
fixed point mentioned above.  In Section III we analyze this unstable
fixed point.  In particular, we find that the RG flow to
this fixed point is governed by two marginally irrelevant operators which
lead to finite-size corrections which only go away with the inverse
logarithm of the chain length.  Fortunately, we are able to calculate
energy eigenvalues for chain lengths up to 5,000 using the Bethe ansatz.
This enables us to analyze in detail the logarithmic behaviour and show
convincingly that our conjecture is correct.  In Section IV we
analyze the effect of perturbing the couplings to the impurity. We
conjecture a general RG flow and phase diagram and attempt to test it
numerically.  The non-integrability  limits the maximum chain length to
about 20.  Because of the presence of two marginal operators it is
difficult to draw definitive conclusions from these calculations but they
seem to be consistent with our conjecture that the
integrable impurity model corresponds to an unstable fixed point.

\section{Review}\label{sec:review}

The continuum limit of the $S=1/2$ Heisenberg antiferromagnet,
$\vec S_i=(1/2)\vec\sigma_i$, can be
written in terms of a free boson, with a particular value of the
``compactification radius'' (or Lagrangian normalization) which enforces
the $SU(2)$ symmetry. (For a review see Ref.~\onlinecite{imp1}.)  This
model is equivalent to the $SU(2)$ Wess-Zumino-Witten model with Kac-Moody
central charge, $k=1$.  The uniform and staggered magnetization
correspond to two different operators. Thus the spin operators,
$\vec S_i=(1/2)\vec\sigma_i$, becomes:
\begin{equation}
\vec S_i \approx
(\vec J_L+\vec J_R) + (-1)^i\hbox{constant}\cdot {\rm tr}h\vec
\sigma.
\label{eq:s}
\end{equation}
Here $\vec J_L$ and $\vec J_R$ are the left and
right-moving spin densities or currents and $h$ is an $SU(2)$ matrix
field.  The current operators have scaling dimension $x=1$ while  $h$ has
$x=1/2$.  Using the operator product expansion, it can be shown that,
\begin{equation} \vec S_i\cdot \vec S_{i+1} \approx
\hbox{constant}(-1)^i\hbox{tr}h. \end{equation}
We now review the effect of local perturbations upon the open and
periodic chain fixed points.

The various types of local
perturbations of the {\em periodic} chain corresponding to quantum impurities
can all be expressed in terms of $\vec J \equiv \vec J_L+\vec J_R$ and
$h$, the former operator being marginal and the latter relevant.  In
fact, most perturbations generate  relevant operators, the only exception
being perturbations symmetric under site-parity ${\cal P}_S$, ie.
reflection about a site, which {\em do not} involve an external spin.
In this
case the parity symmetry ensures that all terms involving $\hbox{tr}h$
cancel.  A perturbation which is not invariant under ${\cal P}_S$ will
generate $\hbox {tr}h$ under a renormalization group transformation.
Let us now consider perturbations that {\em do} involve an external spin.
A ${\cal P}_S$ invariant coupling, i.e. $\vec \sigma_i\cdot\vec S$,
to an external
spin $\vec S$ generates the relevant operator $\hbox{tr}h \vec \sigma
\cdot \vec S$, as can be seen from Eq.~(\ref{eq:s}).
(Note that $\hbox{tr}h \vec \sigma\cdot \vec S$ is
a relevant operator since a decoupled impurity has zero scaling
dimension.)  For equal Heisenberg coupling of two
neighbouring chain-spins to the impurity, symmetric under {\it link}-parity,
i.e. $\vec
\sigma_1\cdot\vec S+\vec \sigma_{L}\cdot\vec S$, the $\hbox{tr}h
\vec \sigma\cdot \vec S$ terms cancel.
However, $\hbox{tr}h$ is generated
since in this case the site-parity, ${\cal P}_S$, is broken.
Hence we arrive at the important
conclusion that a periodic chain with a decoupled impurity is not a
stable fixed point, since relevant operators always will be present.

The situation is different for the {\em open} chain.  In this case  the
boundary operator formalism identifies left and right-moving operators
and the chain-end spins become: \begin{equation} \vec S_{\pm} \propto
\vec J_{\pm}. \end{equation} Here $+$ and $-$ refer to the two sides of
the break in the chain and $\vec J \equiv \vec J_L\equiv J_R$.  Since
$h$ doesn't appear as a boundary operator, all perturbations of the open
chain are, at most, marginal.  A weak coupling of the two sides of the
break is irrelevant.  A weak coupling to an external spin generates
$(\vec J_+ +\vec J_-)\cdot \vec S$, which is analogous to a Kondo
coupling.  It is marginally relevant for antiferromagnetic coupling and
marginally irrelevant for ferromagnetic coupling.  Hence the open chain
with no decoupled impurity or with an impurity
whose coupling flows to zero from the ferromagnetic side are stable
fixed points.

Let us now consider the stable fixed points for open spin-chains with
{\it link}-parity symmetric couplings to an external $S=1$ impurity, a
class of models which includes the integrable one.  The case of a simple
Heisenberg coupling: \begin{equation} H_{\rm imp} = J(\vec \sigma_1+\vec
\sigma_L)\cdot \vec S\label{Heis}\end{equation} was discussed
earlier~\cite{imp1}.  If $J<0$, it renormalizes to zero leaving the open
chain with a decoupled $S=1$ impurity.  If $J>0$ it is marginally
relevant and we assume that it renormalizes to $\infty$.  This produces
an open chain with two sites removed and no leftover impurity in the
low-energy theory.

The excitation spectrum of a long chain of length $L$ contains towers of
states with spacings of $O(1/L)$ up to higher order corrections.  This
low-energy spectrum, which is a universal property of the fixed point, is
reviewed in Ref.~\onlinecite{imp1}, for periodic and open chains.  Some
of the
first few states are given in Table~\ref{spectrum} of the present paper.
Note that spin chains of even length, $L$, with periodic or open boundary
conditions, have parity even (odd) ground-states for $L/2$ even (odd). To
$O(1/L)$, the spectra are identical for $L/2$ even or odd apart from a
parity flip for all states.  Thus we see that in specifying the various
fixed points we must be careful to specify the ground-state parity.  The
infinite antiferromagnetic $J$ fixed point referred to in the previous
paragraph corresponds to a ground-state with reversed parity compared to
$J=0$, since two spins have been effectively removed from the chain to
screen the impurity.  We will take the original chain length to have
$L/2$ even. Thus this screened fixed point has a ground-state with
$S_T^P=0^-$, where $S_T$ is the spin of the state and $P$ its parity.  We
label the renormalization group fixed point corresponding to an open
chain with
 no parity flip as open$^+$;we label the fixed point with the parity flip
as open$^-$.  The open chain with no parity flip and a leftover decoupled
$S=1$ impurity is labeled open$^+ \times (S=1)$.  Thus a negative $J$
renormalizes to the open$^+ \times (S=1)$ fixed point and a positive $J$
renormalizes to the open$^-$ fixed point.

Let us now consider the integrable impurity model,
of Eq. (\ref{eq:hamil}).  The
Bethe ansatz results of Ref.~\onlinecite{schlottmann} indicate that the
magnetization has Curie form as $T \to 0$ with magnitude corresponding to
a decoupled $S=1/2$ impurity.   It is quite easy to see that such a fixed
point will not arise in a link-parity invariant model from the type of
analysis used above, where it is assumed that all couplings renormalize
to $\infty$ or $0$.  If a single chain-spin coupled most strongly to the
impurity it could partially screen it, leaving an effective $S=1/2$
impurity.  However, with  link-parity, this type of analysis always
produces changes in the effective spin by integer units.  The other
possibility is that the $S=1$ impurity effectively ``splits in half'',
donating an extra $S=1/2$ to the chain and leaving behind a decoupled
effective $S=1/2$ impurity. [See Fig.~\ref{fig:split}.] The coupling
of the rest of the chain to the ``donated'' $S=1/2$ must ``heal'', ie.
renormalize to the same value as in the rest of the chain.  Such a
healing phenomena was shown to occur for an $S=1/2$ impurity
coupled symmetrically to two neighbouring sites in a chain~\cite{imp1}.
However,
this does not correspond to a stable fixed point in the present case
because of the decoupled $S=1/2$ impurity. A residual coupling of this to
the healed chain is relevant, as discussed above.  It generates the
operator, $-\lambda'\hbox{tr}h\vec \sigma
\cdot \vec S_{\rm eff}$, where $\vec
S_{\rm eff}$ is the effective $S=1/2$ impurity.

We now propose a  resolution of this dilemma. Due to the
very particular nature of the integrable Hamiltonian,
Eq.~(\ref{eq:hamil}), the relevant
coupling, $\lambda'$, referred to above,
vanishes.  Of course, if we were to make an
infinitesimal change in any of the lattice coupling constants near the
impurity, we should expect that this relevant coupling in the fixed point
Hamiltonian would generally become non-zero.  In the next section we
explore, using both RG  and finite-size Bethe ansatz analysis, this
hypothesis about the integrable impurity model itself.  In Section IV we use
RG and the modified Lanczos method to study perturbations of the
integrable impurity model.

\section{The Integrable Impurity Model}\label{sec:ian}
We now want to study the
integrable impurity
model in more detail.  As stated above, we hypothesize that it
renormalizes to the {\it unstable} fixed point corresponding to the $S=1$
impurity breaking up into two $S=1/2$ spins, one of which is adsorbed
into the chain and the other of which decouples. [See
Fig.~\ref{fig:split}.] We assume that the
chain originally had length $L$; hence, after adsorbing the extra $S=1/2$
variable, it has the effective length $l=L+1$. For this decoupling to occur,
the relevant coupling of the extra spin to the chain, discussed in the
previous section, must be ``fine-tuned'' to zero.  The next most
important coupling to consider is then the marginal coupling of the
impurity to the periodic chain.  This can be written:
\begin{equation}
\delta H = -\lambda(l) v(\vec J_L+\vec J_R)(0)\cdot \vec S_{\rm eff} ,
\label{eq:three}
\end{equation}
where $\vec J_{L,R}(x)$ is the
spin-density of left (right)-movers and $\vec S_{\rm eff}$ is the
effective impurity spin assumed to have size $1/2$.   $v$, the
spin-wave velocity,  plays the role of the velocity of light in the
conformal field theory.
Its value, $v=\pi /2$, is known exactly from the Bethe
ansatz.  A positive $\lambda$ corresponds to a ferromagnetic
coupling.  The total spin of the
left-movers is given by:
\begin{equation} \vec {S}_L= (1/2\pi )\int
dx \vec J_L,
\end{equation}
and similarly for the right-movers.  The total
spin of the periodic chain (not including $\vec S_{\rm eff}$) is
\begin{equation} \vec {S}_{\rm chain} = \vec {S}_L +\vec {S}_R.
\end{equation}

The $\beta$-function for $\lambda$ is calculated in Appendix B of
Ref.~\onlinecite{nuclb}. [See Eq. (B10).   There a positive $\lambda$ is
antiferromagnetic.]  The calculation is identical to that for the Kondo
problem and the result is:
\begin{equation} d\lambda /d \hbox{ln} l =
-\lambda^2.  \end{equation}
Although, in that appendix we only have the left-moving
part of $\vec J$, the two parts of the interaction renormalize
separately so we get the same $\beta$-function.  Note that a
ferromagnetic coupling is marginally irrelevant.
 Solving, we obtain the effective coupling constant at scale $l$ in terms
of the effective coupling constant at scale $l_0$ as:
\begin{equation}
\lambda (l) \approx {\lambda (l_0)\over
1+\lambda(l_0)\ln(l/l_0)}\label{eq:beta}.\end{equation}
Thus, if the integrable impurity model is to renormalize to the proposed fixed
point, the marginal coupling, $\lambda$ must be ferromagnetic (ie.
$\lambda >0$) in addition to the relevant coupling vanishing.  There is
no particular reason for the marginally irrelevant coupling, $\lambda$ to
be strictly zero, and indeed we shall see that it is not.  Such a
marginally irrelevant coupling leads to corrections to the asymptotic
behaviour which only vanish as $1/\hbox{ln} l$.  Consequently, it becomes
difficult to conclude very much about the critical behaviour from
finite-size calculations unless exponentially long chains can be studied.
Fortunately, this is possible using the Bethe ansatz.

A similar difficulty was already encountered for the periodic $S=1/2$
chain, without any impurity.  In that system there is a ``bulk'' marginal
operator:
\begin{equation} \delta H = -g(l)v(8\pi^2 /\sqrt{3})\int dx \vec
J_L(x)\cdot \vec J_R(x). \end{equation}
(Recall that dimension {\it two}
bulk operators are marginal, but dimension {\it one} boundary operators
are marginal; the difference arises from the integral over $dx$ in the
former case.) In this case, the renormalized coupling is given
by~\cite{jphysa}:
\begin{equation} g(l)=g(l_0)/(1+4\pi g(l_0)\ln(l/l_0)/\sqrt{3}).
\label{betag} \end{equation}
As first pointed out by Cardy~\cite{cardy86},
and applied to the study of periodic
Heisenberg chains in Ref.~\onlinecite{jphysa}, the effect of such marginally
irrelevant couplings on the finite-size spectrum can be calculated in
perturbation theory in the effective coupling constant.

At a conformally invariant fixed point, excitation energies, take the
form~\cite{cardy84}:
\begin{equation}
E_n-E_0 = {2\pi v\over l}x_n,
\end{equation}
where
$E_0$ is the ground-state energy and the scaling dimensions, $x_n$, are
universal.  The $x_n$'s for the lowest energy state of given ${
S}_L$, ${S}_R$ can be written:
\begin{equation}
x_n = ({
S}_L)^2+({S}_R)^2 \label{x_n}.
\end{equation}
[See for example Ref.~\onlinecite{jphysa}.]
The total spin multiplets are determined by the usual angular momentum
addition rules, $|S_L-S_R| \leq S_{\rm chain} \leq S_L+S_R$.
For an even length chain  ${S}_L$ and ${S}_R$ are either
both integer or both half-integer; for an odd length chain one of them is
integer and one is half-integer.

The ground-state energy takes the form:
\begin{equation}
E_0 = \epsilon_0 l -{\pi vc\over 6l},
\label{eq:seventeen} \end{equation}
where $\epsilon_0=\ln 2$, the ground-state
energy density, is non-universal and the universal $1/l$ correction is
proportional to $c$, the conformal anomaly parameter; $c=1$ for the
$S=1/2$ chain.

The excitation energies receive corrections in first order perturbation
theory in the bulk marginal coupling constant, $g$~\cite{cardy86,jphysa}:
\begin{equation}
\delta x_n = -{4\pi \over \sqrt{3}} g(l) \vec {S}_L\cdot \vec
{S}_R.\end{equation}
Note that this dot product can be determined
from $S_L$, $S_R$ and $S_{\rm chain}$:
\begin{eqnarray}
{S}_L\cdot \vec
{S}_R& =& (1/2)[(\vec {S}_L+\vec {S}_R)^2 -  \vec {
S}_L^2 - \vec {
S}_R^2]\nonumber\\
&=&(1/2)[S_{\rm chain}(S_{\rm chain}+1)-S_L(S_L+1)-S_R(S_R+1)].
\end{eqnarray}
The ground-state energy only obtains a correction of third order in
$g(l)$~\cite{cardy86,jphysa}:
\begin{equation}
\delta c = {2\pi g^3(l)\over \sqrt{3}}.
\end{equation}

The integrable impurity model has two marginal coupling constants, $g(l)$
and $\lambda (l)$, producing two sources of logarithmically slow
finite-size behaviour.  The corrections due to $g$ will be the same as for
the periodic chain, given above.  We now calculate the corrections due to
the marginal boundary coupling constant, $\lambda$.

There is a first order correction to the excitation energies.  Since, for
$\lambda=0$, the chain is translationally invariant, for any eigenstate,
\begin{equation} <n| (\vec J_L+\vec J_R)(0)|n>=[2\pi /l]<n|\vec S_{
\rm chain}|n>. \end{equation}
The marginal coupling of the impurity to the periodic chain,
Eq.~(\ref{eq:three}), will then give rise to a finite size correction of
the scaling dimension of the following form
\begin{equation}
\delta x_n =
-\lambda(l)\vec {S}_{\rm chain}\cdot \vec S_{\rm eff}.
\end{equation} This can be expressed in terms of the observable total
spin of the state, $S_T$ where $\vec S_T= \vec{S}_{\rm chain}+\vec
S_{\rm eff}$, the spin of the periodic chain, ${S}_{\rm chain}$, and the
impurity spin, $S_{\rm eff}=1/2$, giving: \begin{equation} \vec {
S}_{\rm chain}\cdot \vec S_{\rm eff}=(1/2)[S_T(S_T+1) - {S}_{\rm chain}(
{S}_{\rm chain}+1)-3/4].\end{equation}  Combining the various terms we
obtain
\begin{equation}
E_n-E_{1/2} = (2\pi v/l)\left[S_L^2+S_R^2 -{4\pi
\over \sqrt{3}} g(l) \vec {S}_L\cdot \vec {S}_R-\lambda
(l)\vec {S}_{\rm chain}\cdot \vec S_{\rm  eff}\right],
\label{eq:ex}
\end{equation}
where the effective length, $l$, for the integrable impurity model is $l=L+1$.

The correction to the  energy of the ground-state, which has
$S_L=S_R=S_{\rm chain}=0$, $S_T=1/2$, vanishes to first order in $\lambda$,
so let us consider next order.  We can express the second order
correction in terms of the expansion of the partition function in powers
of $\lambda$ in the zero-temperature limit.  This gives:
\begin{equation}
Z=Z_0[1+(1/2)\int_{-\infty}^{\infty} d\tau_1d\tau_2<\delta H(\tau_1)\delta
H(\tau_2)>+...]
\label{eq:tintegral}
\end{equation}
Thus the correction to the ground-state
energy is:
\begin{equation} \delta E_0 = -(1/2) (v\lambda
)^2\int_{-\infty}^{\infty} d\tau  <0|(\vec J_L+\vec J_R)(0,0)\cdot \vec
S_{\rm eff}(0) (\vec J_L+\vec J_R)(\tau ,0)\cdot \vec S_{\rm eff}(\tau
)0>.
\end{equation}
In lowest order perturbation theory:
\begin{equation}
<0|S^a_{\rm eff}(0)S^b_{\rm eff}(\tau )|0>=(1/3)\delta ^{ab}\vec S_{
\rm eff}^2= (1/3)\delta ^{ab}s(s+1)=(1/4)\delta ^{ab}.
\end{equation}
We also
need:
\begin{equation} <0|J^a_L(\tau )J^b_L(0 )|0>= \delta^{ab}/2(v \tau
)^2,
\end{equation}
and the same for $J_R$.  (Left and right are
uncorrelated.  See Ref.~\onlinecite{nuclb}  for the normalization.)  This
is just the free fermion current Green's function.  This is the result
for an infinite system.  To get the result for a finite system we make a
conformal transformation to map the infinite plane onto the cylinder of
circumference $l$, or else just work out explicitly the free fermion
current Green's function with appropriate boundary conditions.  The
result is:
\begin{equation}
<0|J^a_L(\tau )J^b_L(0 )|0>= \delta^{ab}
{1\over 2[(l/\pi )\sinh (\pi v\tau /l)]^2}.
\end{equation} Note that
the $\tau$ integral in Eq.~(\ref{eq:tintegral})
diverges as $\tau \to 0$.  This is an ultraviolet
divergence which would be cut off by the lattice spacing of the spin
chain.  It is simplest just to put in a cut off on the $\tau$ integral,
$|\tau |>\tau_0$.  To evaluate the integral we change variables to
$u=\hbox{tanh}(\pi v\tau /l)$, giving:
\begin{equation} \delta
E_0=-(3/4)\lambda^2v {\pi \over l}\int_{u_0}^1du/u^2,
\end{equation}
where $u_0\approx \pi v\tau_0/l$. (We assume $l>>\tau_0$.)
Thus
\begin{eqnarray}
\delta E_0&=&-(3/4)\lambda^2 v{\pi \over
l}(1/u_0-1)\nonumber \\ &=&-3 \lambda^2/4\tau_0 +  3 \lambda^2 v(\pi
/4l).\end{eqnarray}
Note that the first, cut-off dependent term, is a
non-universal contribution to the $l$-independent part of $E_0$.  The
second is a correction to $c$:
\begin{equation} \delta c = -{9 \over
2}\lambda^2(l).  \label{eq:deltac} \end{equation}
The
universal ground-state energy correction is second order in $\lambda$ but
third order in $g$; we find that the $\lambda$ correction is much
larger. Assembling the various terms the ground-state energy takes the
form:
\begin{eqnarray}
E_{1/2}&-&\varepsilon_0L-\varepsilon_1= -\frac{2\pi
v}{12l}\Big[1+\frac{[2\pi
g(l)]^3}{\sqrt{3}}-\frac{9}{2}\lambda^2(l)\Big],
\label{eq:gs}
\end{eqnarray}
where we should use $l=L+1$ for the integrable impurity model.
Here $\varepsilon_0=\ln 2$, and  from the work of
Schlottmann~\cite{schlottmann}
$\varepsilon_1=(1/2)(\psi(3/4)-\psi(5/4))\simeq -0.4292036733$
where $\psi$ is the digamma function.

The ground-state, with $S_T=1/2$, occurs for $L$ odd corresponding to
the effective length, $l=L+1$, being even.  All states occurring
for even $L$ are regarded as excited states.  In applying Eq.
(\ref{eq:ex}), the values of $g(l)$ and $\lambda (l)$ are expected to
interpolate smoothly between even and odd $l$.
 In table (\ref{tab:states}), we give all the relevant quantum numbers
for the first few lowest energy states.  Note that the states with
half-integer ${S}_{\rm chain}$ all come in pairs of opposite parity,
obtained by interchanging the quantum numbers ${S}_L$ and ${
S}_R$.  These pairs are degenerate including $O(\lambda)$ corrections.
Presumably they are, for more general models,
split by corrections of higher order in
irrelevant operators. However, as we shall see, for the integrable
impurity model they remain exactly degenerate.

In what follows, we test these formulas in two different ways.  One
way is to confirm that all energy levels that we consider are given by
these formulas with the same values for $g(l)$ and $\lambda (l)$ for
a given length $l=L+1$.  We expect small discrepancies to occur  because of
corrections in higher orders of perturbation theory; however, these
should become smaller at larger $l$.  Secondly, the functions $g(l)$ and
$\lambda (l)$ should be given by the lowest-order $\beta$-function
results, Eqs. (\ref{eq:beta}) and (\ref{betag}) for sufficiently large
$l$.

Following Andrei and Johannesson~\cite{andrei} the Bethe ansatz
equations, for the integrable Hamiltonian of Eq.~(\ref{eq:hamil}), are
\begin{equation}
\left(\frac{\Lambda_k+i1/2}{\Lambda_k-i1/2}\right)^{L}
\left(\frac{\Lambda_k+i}{\Lambda_k-i}\right)
=-\prod_{l=1}^M\frac{\Lambda_k-\Lambda_l+i}{\Lambda_k-\Lambda_l-i},
\label{eq:bae}
\end{equation}
where $L$ is the number of  $S=1/2$'s.
The number of roots, $M$ determines the total $S^z$ component through
the relation $S^z=L/2+1-M$.
In
terms of the solutions, $\Lambda_k$, to the Bethe ansatz equations,
Eq.~(\ref{eq:bae}), the energy is given by \begin{equation}
E=-\sum_{k=1}^M\frac{1/2}{\Lambda_k^2+(1/2)^2}. \end{equation} One should
note that the energy, $E_H$, of the Hamiltonian Eq.~(\ref{eq:hamil}) is
related to $E$ by $E_H=E+L/4+2/9$.

The Bethe ansatz equations are solved numerically by first making the
assumption that the solutions occur in strings of length $n$.
\begin{equation}
\Lambda_j^{n,\alpha}=\Lambda_j^n+i(n+1-2\alpha)/2,\,\,\alpha=1,\dots ,n,
\end{equation}
and then solving Eq.~(\ref{eq:bae}) for the centers of the strings,
$\Lambda_j^n$. If $\nu_n$ is the number of strings of length $n$ then we
must have
\begin{equation}
M=\sum_n n\nu_n.
\end{equation}
The string
assumption is then relaxed and the full Bethe ansatz equations are solved
by a Newton-Raphson method using the string solution as the starting
point.

For chains of even length, $L+1$, corresponding to an {\it odd} number,
$L$, of sites with $S=1/2$ and one spin $S=1$, we determine the three
lowest lying levels, $E_{1/2}, E_{3/2}$ and $E_{1/2}^*$. Here the index
refers to the total spin, $S_T$, of the state. The ground-state
$E_{1/2}$ corresponds to a
solution with $(L+1)/2$ real roots and no strings, $E_{3/2}$ is a solution
with $(L-1)/2$ real roots and $E_{1/2}^*$ has $(L-3)/2$ real roots plus a
2-string at $x\pm i(\pi/2+\delta)$ where $\delta$ is a small positive
number quickly approaching zero for long chains, and $x$ is non-zero. The
results for these three levels are given in Table~\ref{tab:even}. For
chains of odd length, $L+1$, corresponding to an {\it even} number, $L$,
of sites with $S=1/2$ plus one site with spin $S=1$, we determine the
two levels $E_1$ and $E_2$. $E_1$, the lowest lying
state, is a solution of the Bethe ansatz equations with $L/2$ real roots.
We were not able to obtain the first excited state, $E_0$,
by the Bethe ansatz scheme.
The next excited level, $E_2$, has $L/2-1$ real
roots. The results for $E_1$ and $E_2$ are summarized in
Table~\ref{tab:odd}.

These states exhibit  some remarkable degeneracies, for finite $L$. Eq.
(\ref{eq:bae}) is invariant under the operation, $\{\Lambda_k\} \to
\{-\Lambda_k\}$.  Thus, in cases where the set of roots is not symmetric
about $0$, a pair of degenerate solutions is obtained, if we assume
that the corresponding wave-functions are linearly independent as can be
verified for short chains.
This is the case
for all solutions discussed above with even $L$ and also for $(1/2)^*$.
The even $L$ degeneracies can be understood from the picture of the RG
fixed point corresponding to a periodic chain of $(L+1)$ $S=1/2$'s and a
decoupled $S=1/2$.  The periodic odd-length chain has a large exact
degeneracy for finite $L$.  This follows from the fact that the
ground-state does not have zero momentum.  Instead it consists of two
degenerate doublets, $(1/2)^{\pm}$ with momentum $\pm k_0$.  As $L \to
\infty$, we expect $k_0 \to \pi /2$.  By forming linear combinations of
these states we can form positive and negative parity eigenstates.  All
low-lying states have momentum close to $\pm \pi /2$ and consequently
also come in parity doublets.  In the conformal field theory picture, the
parity doubling arises from the fact that the periodic chain of odd
length, $L+1$ has $S_L$ integer and $S_R$ half-integer or vice versa.  The
corrections to the excitation energies of first order in $g$ and
$\lambda$, Eq. (\ref{eq:ex}) do not lift the degeneracy.  This must be
true to all orders in $g$ and all other irrelevant bulk operators for the
periodic chain.  On the other hand, we expect that higher order
corrections in irrelevant boundary operators will, in general, lift the
degeneracy since momentum is not, in general, a well-defined quantum
number for the impurity system.  Remarkably, this does not happen
for the integrable impurity model and the
degeneracy remains exact.  Even more surprising is the degeneracy for odd
$L$.  In this case the corresponding periodic chain has even length,
$L+1$ and does not exhibit any exact finite length degeneracies.
Nonetheless, such a degeneracy occurs for the $(1/2)^{\pm *}$ states.  We see
from Table~\ref{tab:states} that this occurs via a cancellation of the
$O(g)$ and $O(\lambda )$ marginal corrections.  At large $l$,
using $\lambda(l)\rightarrow 1/\ln(l)$, $g(l)\rightarrow \sqrt{3}/4\pi\ln(l)$
the correction to $x_n$ becomes:
\begin{equation} \delta x_n = -[\vec
S_L\cdot \vec S_R +\vec S_{\rm chain}\cdot \vec S_{
\rm eff}]/\hbox{ln}l.\end{equation}
$[\vec S_L\cdot \vec S_R +\vec
S_{\rm chain}\cdot \vec S_{\rm eff}]$ has the value $-3/4$ for both the
$(1/2)^{*+}$ and $(1/2)^{*-}$ state.  The splitting of these levels by
the marginal bulk operator is cancelled by the splitting due to the
marginal boundary operator! Furthermore, this cancellation must be exact
to all orders in irrelevant bulk and boundary operators.  This suggests
the existence of some sort of hidden symmetry in the integrable impurity
model reminiscent of the Yangian symmetry discussed recently for the
$1/r^2$ periodic Heisenberg chain~\cite{Yangian}.

%A direct of way of testing the conjecture that the fixed point
%corresponds to a
%chain with an extra spin and a decoupled $S=1/2$  is to measure the
%scaled energy gap, $(E_0-E_1)(L+1)/\pi v$.  We see from Table
%(\ref{spectrum}) or (\ref{tab:states}) that this gap should  converge to
%$0$ for odd chain lengths, $L+1$.
% Fig.~\ref{fig:gap}  shows the scaled energy gap, $(E_0-E_1)L/\pi v$, for
%odd chains of length up to $L=84$. Clearly our results are consistent with
%the spectrum of an odd  chain. Since we were not able to determine $E_0$
%for chains longer than 84 we shall hereafter not discuss it.
%
We can now try to extract the coupling constant, $\lambda(l)$ defined in
Eq.~(\ref{eq:three}), as determined from the five levels, $E_{1/2},
E_{3/2}$, $E_{1/2}^{*-}$, $E_1$ and $E_2$.
The bulk marginal coupling constant, $g(l)$, has
already been determined from finite-size analysis of long {\it periodic}
chains without the impurity~\cite{jphysa}.
We shall use this as input and determine the boundary marginal
coupling, $\lambda (l)$ from our data on the integrable
impurity model.  Thus the
only free parameter is $\lambda (l)$.  Note that $(1/2)$ is the
ground-state so we fit its energy to Eq. (\ref{eq:gs}).  The other four
are excited states so we fit their excitation energies by Eq.
(\ref{eq:ex}).  The different estimates of $\lambda (l)$ obtained from
the different energy levels are shown in Fig.~\ref{fig:couplings}.
Notice, first of all that $\lambda$ is positive, corresponding to a
ferromagnetic, marginally irrelevant coupling of $\vec S_{\rm eff}$ to
the chain. As expected the different estimates of $\lambda (l)$ are all
approximately the same.  As $l$ increases, the estimated couplings
approach each other and get smaller.

We compare $\lambda(l)$ to the one loop $\beta$ function result
Eq.~(\ref{eq:beta}), $\lambda_{\rm rg}$, in Fig.~\ref{fig:beta}. It is
seen that the first-order $\beta$-function result is valid for chains
longer than $l\sim 100$, indicating that our perturbative results should
be meaningful for chains of this length or longer. A similar plot for the
other coupling $g$ is shown in Fig.~\ref{fig:betag}. Again  we see that
the one loop $\beta$ function gives good results for chains longer than a
few hundred. As shown in Table~\ref{tab:states} the excited state,
$(1/2)^{*+}$, does not receive a correction to its excitation energy of
first order in
$\lambda$. We can therefore use it to extract an estimate of $g(l)$
independent of the results from Ref.~\onlinecite{jphysa}. This estimate
, $g_{1/2}(l)$, is also shown
in Fig.~\ref{fig:betag}. It compares nicely to the $\beta$ function
results as well as to the results obtained for the pure periodic $S=1/2$
chain in Ref.~\onlinecite{jphysa}. [The corresponding plot in
Ref.~\onlinecite{jphysa} contained a numerical error in the calculation
of $g_{\rm rg}$.]

{}From the above results we conclude that the spectrum of the integrable
impurity model indeed asymptotically becomes that of a periodic
chain with $L+1$ spin-1/2's and a decoupled $S_{\rm eff}=1/2$.
The impurity spin $S=1$ has effectively split in half. The spectrum
corresponds to column 1 in Table~\ref{spectrum}.

\section{General Hamiltonians}
In this section, we consider the effects
of perturbing the
integrable impurity Hamiltonian $H_{\rm Int}$, Eq.~(\ref{eq:hamil}),
away from its integrable form.  A general
phase diagram is hypothesized and is tested using the modified Lanczos
method for general Hamiltonians on chains with $L \leq 20$. To be
concrete, we consider the following two-parameter set of Hamiltonians:
\begin{eqnarray}
H&=&\frac{1}{4}\sum_{i=1}^{L-1}{\vec \sigma}_i\cdot{\vec
\sigma}_{i+1} + H_I\nonumber\\
H_I&=&{ 2J_1 \over 9}[(\vec \sigma_1+\vec \sigma_L)\cdot \vec S
+ {1\over 2}\{\vec \sigma_1\cdot \vec S , \vec \sigma_L\cdot \vec
S\}-{7\over 8}\vec \sigma_1\cdot \vec \sigma_L]+{J_2\over 2}(\vec
\sigma_1+\vec \sigma_L)\cdot \vec S. \label{Hamgen}
\end{eqnarray}
Here $\vec S$ is the $S=1$ impurity.
$J_1=1$, $J_2=0$ is the integrable impurity model,
and $J_1=0$ corresponds to the models discussed in Ref.~\onlinecite{imp1}.
$J_1=J_2=0$ is the simple open chain with a
decoupled $S=1$ impurity.  By combining our understanding of the phase
diagram near $|J_i|=0$, $|J_i| \to \infty$
and the integrable impurity model, we
hypothesize a general phase diagram.

The vicinity of $|J_i|=0$ can be readily analyzed.  Near the open chain
fixed point the terms
$\{\vec \sigma_1\cdot \vec S , \vec \sigma_L\cdot \vec S\}$ and
$\vec \sigma_1\cdot \vec \sigma_L$ in Eq. (\ref{Hamgen})
contain a product of {\it both} chain-end spins $\vec \sigma_1$ and $\vec
\sigma_L$. In the continuum limit they correspond to a dimension 2
irrelevant operator containing the product of $\vec J_+$ and $\vec J_-$.
Thus for small
$J_i$ we may approximate the impurity part of the Hamiltonian,
$H_I$, by $(2J_1/9 +J_2/2)(\vec
\sigma_0+\vec \sigma_L)\cdot \vec S$.
This is marginally irrelevant for
$(J_2+4J_1/9)<0$.  Thus, for $(J_2+4J_1/9)<0$, the system renormalizes to the
fixed point consisting of the open chain with a decoupled $S=1$ impurity
which we denote by open$^+ \times (S=1)$.
For $(J_2+4J_1/9)>0$ the coupling is {\it antiferromagnetic} and
therefore marginally relevant. We expect it to
renormalize to $\infty$, screening the impurity and effectively removing
two sites from the chain.
This produces a parity flip, giving the open$^-$ fixed point.
(Recall from Sec. II that the removal of
two spins from the open chain by the screening process reverses the
parity.) This transition is $\infty-$order since it is driven by
a marginal operator, i.e. the cross-over length scale diverges
exponentially as $(2J_1/9 +J_2/2)\rightarrow 0$ from positive values.

Let us next consider what happens in the limit $|J_i|\to \infty$.  In
this limit we must find the ground-state of the
three-spin system,  $\vec S$, $\vec \sigma_1$, $\vec \sigma_L$,
of $H_I$.
Depending on the
$J_i$'s, the ground-state has spin and parity $0^+$, $1^+$ or $2^+$.
These wave-functions are depicted schematically in Fig.~\ref{fig:3spin},
as are the regions of stability of the three states.
The $0^+$ case leads directly to a stable fixed
point when we include the relatively weak coupling of these three spins
to the rest of the chain.  These couplings are irrelevant, leaving the
open chain fixed point open$^-$.
The $2^+$ impurity state, however, is unstable because the effective
$S=2$ impurity is coupled antiferromagnetically to the chain.  Assuming
this coupling flows to $\infty$, the impurity is partially screened,
leaving an effective $S=1$ impurity with an a {\it ferromagnetic}
coupling to the rest
of the chain. Hence, it will renormalize to $0$.
Since four sites are involved in producing the effective $S=1$ decoupled
impurity, there is no parity flip.  Thus we obtain the open$^+\times
(S=1)$  phase.
Of special interest is the $1^+$ ground-state for the three-spin complex.
$\vec \sigma_1$ and  $\vec \sigma_L$ form an $S=1$ spin, symmetric under
interchanging these two spins.  This then couples to the $S=1$ impurity
to form a state of total spin $1$.
Note that
neither the minimal spin
(0) nor the maximal spin (2) occurs.
The biquadratic term in $H_I$, involving the
anticommutator, is necessary to assure that this happens.
%This  is possible only if a
%biquadratic term is added to the Hamiltonian.
%otherwise one of the other
%two states of the three-spin system is always the ground-state.
Thus we
begin to see how the integrable impurity model
can have different behaviour than
the simpler one discussed above, with $J_1=0$.  The $1^+$
state of the three-spin cluster does not correspond to a stable fixed
point since the rest of the chain is coupled antiferromagnetically to the
effective $S=1$ impurity, and such a coupling is marginally relevant.  We
expect this coupling to the next two spins in the chain, $\vec \sigma_2$
and $\vec \sigma_{L-1}$ to also renormalize to $\infty$, screening the
effective impurity and leading to an open chain fixed point with no
leftover impurity spin.  In this case, four chain-spins are involved in
screening the impurity in a parity-symmetric way, so that the stable
fixed point is open$^+$.

For large $|J_i|$, we expect the above three different stable phases to
occur, with the phase boundaries asymptotically approaching those of the
three-spin system, as drawn in Fig.~\ref{fig:phase}.  Note that the
open$^+\times (S=1)$ phase is equivalent to $J_i=0$, an open chain with a
decoupled $S=1$ impurity; we may think of this entire phase as
renormalizing to the origin.  The open$^+$ and open$^-$ phases, however,
are characterized by impurity couplings renormalizing to $\infty$.

We have established, in the previous section, that the integrable impurity
model
renormalizes to a fixed point in which the $S=1$ impurity effectively
splits in half, one extra $S=1/2$ being absorbed by the chain and the
other decoupling. This can only occur if no relevant operator
connects the periodic chain to the decoupled $S=1/2$ spin. We now analyze
the effect of small perturbations around the integrable impurity
model by
considering the periodic chain with both marginal and relevant couplings
to the effective $S=1/2$ impurity, i.e.
we consider the continuum limit Hamiltonian consisting of the $k=1$
WZW model with two local perturbations:
\begin{equation} \delta {\cal H}
= -[\lambda^{'}\hbox{tr}h\vec \sigma +\lambda (\vec J_L+\vec J_R)]\cdot
\vec S_{\rm eff}.
\label{eq:4.2}
\end{equation}
The integrable impurity
model has $\lambda^{'}=0$,
$\lambda >0$; ie. the relevant coupling vanishes and the marginal one has
the irrelevant sign.  An infinitesimal perturbation of the integrable
impurity
model will, in general, produce a non-zero $\lambda^{'}$.  The resulting
behaviour was discussed in Ref.~\onlinecite{imp1}, in the context of a
simple coupling of an $S=1/2$ impurity to a single site in a periodic
chain.  We expect that $\lambda^{'}$ will renormalize to $\pm \infty$
beginning from an infinitesimal positive (or negative) value. The
negative  case, corresponding to an antiferromagnetic coupling, leads to
screening of  $\vec S_{\rm eff}$ by a single site in the chain.  The
stable fixed point is an open chain with one site removed, open$^+$. For
$\lambda^{'}>0$, the ferromagnetic case, $\vec S_{\rm eff}$ and the site
to which it is coupled form an effective $S=1$ impurity.  However, this
is not a stable fixed point.  The $S=1$ effective impurity is coupled
antiferromagnetically to two neighbouring spins.  We expect this
coupling to renormalize to $\infty$, screening the effective impurity.
Once again the stable fixed point is an open chain.  However, in this
case, {\it three} chain spins are involved in the screening process and
get removed from the effective open chain at the stable fixed point.  As
discussed in Section II, the parity of all low-energy states is flipped
relative to the case where a single chain-spin is removed.  Thus the
stable fixed point in this case is the open$^-$.
We see that the unstable critical point, $\lambda^{'}=\lambda =0$,
to which the integrable impurity model
renormalizes separates the stable open$^+$
and open$^-$ phases.

We now turn to a discussion of the order of the phase transition
separating the open$^+$
and open$^-$ phases.
We expect that a second order critical line will
exist for a finite range of positive $\lambda$ with $\lambda^{'}=0$
governed by the $\lambda=\lambda'=0$ critical point. On
the other hand, if the marginal coupling $\lambda<0$, then it is
relevant and renormalizes to large values.  In this case, the simplest
assumption is that $\lambda \to -\infty$; otherwise we would be forced to
postulate another non-trivial critical point.  In general, when impurity
couplings renormalize to $\infty$ we expect a first-order phase
transition.  The reason is that we can then ignore any couplings of the
impurity complex to the rest of the chain.  In this particular case we
can consider only the $S=1/2$ impurity and three chain spins.  This
cluster of four $S=1/2$'s has a $0^+$ or $0^-$ ground-state depending on
the various couplings.  The phase transition in this limit is a simple
level-crossing in the four-spin system and is therefore first order.
The critical
point, $P_1$, at $\lambda=\lambda^{'}=0$ is on the phase boundary between
open$^+$ and open$^-$ phases and separates the second from first-order
transition lines.  The integrable impurity model, which was shown in the last
section to have a non-zero positive $\lambda$, lies on the second-order
part of the phase boundary.

Now we attempt to combine our information about the small $J_i$, large
$J_i$ regions and the vicinity of the integrable point. Our large $J_i$
analysis tells us that there are three stable phases, open$^-$,
open$^+$ and open$^+\times (S=1)$. Our analysis of the vicinity of the
integrable impurity model tells us that it should be on the open$^- -
\hbox{open}^+$ phase boundary.  It renormalizes to a critical point,
$P_1$ in Fig.~\ref{fig:phase} where this transition changes from first
to second order.

The open$^-$ phase occurs when we {\it increase} $J_2$ from
$P_1$.  This is to be expected because $J_2$ corresponds to a coupling
of $\vec S_{\rm eff}$ to the two nearest neighbours, $\vec \sigma_1$, $\vec
\sigma_L$, not to the adjacent spin $\vec \sigma_0$, as shown in
Fig.~\ref{fig:nns}.  Antiferromagnetic $J_2$, i.e. $\lambda'>0$,
leads to a screening of
$\vec S_{\rm eff}$ by $\vec \sigma_0$, $\vec \sigma_1$ and $\vec \sigma_L$.
The removal of $\vec \sigma_0$, $\vec \sigma_1$ and $\vec \sigma_L$
from the open chain leaves a chain with $L-2$ sites and thereby
implies a parity flip. In Fig.~\ref{fig:numJ2} we show a Lanczos
calculation of some low-lying states for length 20, as a function of
$J_2$. The open$^-$ spectrum shown in Table~\ref{spectrum} occurs
for positive $J_2$.
The Lanczos results are discussed in more detail below.

We see that there
must be another multi-critical point in the phase diagram where all three
stable phases meet, $P_2$ in Fig.~\ref{fig:phase}. This point is
presumably not at the origin since
there is only one marginal operator
in the vicinity of the origin, $(2J_1/9 +J_2/2)(\vec
\sigma_0+\vec \sigma_L)\cdot \vec S$, as discussed above,
so we only expect two phases to meet at that point.  We hypothesize
that this higher multi-critical point
corresponds to an open chain with {\it two} $S=1/2$ effective impurities,
$\vec S_1$ and $\vec S_2$, decoupled from the chain and from each other.
ie. the original $S=1$ impurity effectively breaks up into two $S=1/2$
impurities, with everything decoupled at the multi-critical point. There is
one relevant coupling and two marginal ones at this critical point.  We
write these, in the lattice model as:
\begin{equation} \delta {\cal
H} = e\vec S_1\cdot \vec S_2+e_1\vec S_1\cdot (\vec \sigma_1 +\vec
\sigma_L) + e_2\vec S_2\cdot (\vec \sigma_1+\vec \sigma_L).
\label{eq:4.3}
\end{equation}
[See Fig.~\ref{fig:P2}.] We now analyze the phase diagram for this
model.  Since the impurity spins have zero scaling dimension, the
coupling $e$ is highly relevant.  Assuming that a non-zero $e$
renormalizes to $\pm \infty$, the two impurities lock into a singlet
leaving the open chain fixed point, open$^+$, for $e>0$.  For $e<0$ they
lock into an effective $S=1$ impurity.  If both couplings $e_i<0$
(ferromagnetic), this $S=1$ impurity decouples, giving the open$^+\times
(S=1)$ phase.  On the other hand if at least one of the couplings $e_i$
is
antiferromagnetic, then the effective $S=1$ impurity is screened, giving
the open$^-$ phase.  The open$^+\times (S=1) - \hbox{open}^-$ phase
transition is equivalent to the one discussed in the second paragraph of
this section. It is therefore of $\infty$-order.
Note that the open$^+\times (S=1)$ to
open$^+$ phase transition is controlled by the multi-critical point,
$P_2$ at
$e=e_i=0$.  This transition is governed by the relevant coupling constant
$e$, which only involves the two impurities and not the rest of the
chain.  Thus it corresponds to a simple level-crossing in this two-spin
system and so should be first order.  We hypothesize that, as we move
along the critical line, where $e=0$, one of the marginal couplings, say
$e_1$, changes sign at the multi-critical point, while the other remains
ferromagnetic. The transition between open$^-$ and open$^+$ phases,
for $e_1 >0 $, is
governed by the behaviour at  $e_1$ of $O(1)$.  We know that a single
$S=1/2$ impurity with such a coupling to an open chain will get absorbed
by the chain, ie. the defect heals and the fixed point is the periodic
chain with an extra spin.  Thus it is plausible that this phase boundary
is second-order and renormalizes to a periodic chain with a single
decoupled $S=1/2$ impurity. Note that the    continuation of the
open$^+\times (S=1) - \hbox{open}^+$ phase boundary is the open$^- -
\hbox{open}^+$ phase boundary, but the order of the transition changes
from first to second, as drawn in Fig.~\ref{fig:phase}.  This follows
from the fact that both phase boundaries are governed by the vanishing of
the relevant coupling constant, $e$.  The system renormalizes to $P_1$ by
one of the impurities being absorbed into the chain.  It renormalizes to
the origin by the two impurities locking into a decoupled $S=1$ impurity.

Altogether there are five different critical
points; three occurring at finite coupling, $P_1, P_2,$
open$^+\times(S=1)$ and two at infinity, open$^+$, open$^-$, as
shown in Fig.~\ref{fig:phase}.  Various sections of the transition
lines are first, second or infinite order. The
detailed shape of the phase boundaries depicted schematically in
Fig.~\ref{fig:phase} is not known.  What is known is (i) the asymptotic slope
of the three phase boundaries at $|J_i| \to \infty$, (ii) the slope of the
phase boundary at the origin and (iii) the fact that the integrable impurity
model at
$J_1=1$, $J_2=0$ lies on the open$^- - \hbox{open}^+$ phase boundary.

We now discuss our numerical results on chains of length $L \leq 20$.  We
emphasize at the outset that we are fighting finite-size corrections that
vanish logarithmically slowly from two sources: the bulk marginal
coupling, $g$ of Section III which is present everywhere in the phase
diagram and the marginal boundary operator $\lambda$ which is present in
some parts of the phase diagram.  Thus we can only expect our results to
be of anecdotal value.

%Let us first consider the effect of varying $J_1$
%with $J_2=0$, shown in Fig.~\ref{fig:numJ1}.  We present the scaled
%energy gaps, $L\Delta E/\pi v$, for the four states with quantum number
%$0^{\pm}$ and $1^{\pm}$.
In Fig.~\ref{fig:numJ1} we present the scaled
energy gaps, $L(E-E_{1^+})/2\pi v$, for the four states with quantum number
$0^{\pm}$ and $1^{\pm}$ as a function of $J_1$ with $J_2$ fixed at
$J_2=0$. In Fig.~\ref{fig:numJ1} the integrable impurity model
thus corresponds to $J_1=1$.
%At the integrable impurity model, $J_1=0$, all four
%levels are asymptotically degenerate as shown convincingly in Sec. III;
%for finite chains they are split by corrections of $O(1/\hbox{ln}L)$.  We
%see that, for $L=20$, these splittings are sizable.  Note the exact
%degeneracy of the $0^{\pm}$ and $1^{\pm}$ states even for finite $L$ that
%was discussed in Section III.

First, let us consider what happens as we increase $J_1$ away from its value,
$J_1=1$ at the integrable impurity model.  We see that the $0^+- 1^-$ gap drops
rapidly with increasing $J_1$.  This is to be expected since, according to
Fig.~\ref{fig:phase} the system is in the open$^+$ phase.  Note
however, that the $0^+$ state is not the ground-state, even for
$J_1\approx 4$, for $L\leq 20$.  We do expect that it would become the
ground-state for sufficiently large $L$, for any $J_1>1$.  In
Fig.~\ref{fig:numJ1} we show results for two different chain lengths
$L=8$ and $L=20$. As can be seen there is some
evidence that the $1^- - 0^+$ gap indeed is closing with
increasing $L$.
The scaled gaps between the $0^\pm-1^\pm$ states become
asymptotically degenerate at $J_1=1$, the integrable impurity model.
As seen in Fig.~\ref{fig:numJ1} the $0^+$ level has the most negative
slope as a function of $J_1$.
It is then
plausible that as the gap closes at $J_1=1$,
the $0^+$ state cross the $1^-$ state
for $J_1>1$.
Even for large $J_1$ this level-crossing only takes place at large $L$.
The reason is that as $J_1\to \infty$ for fixed
$L$, we obtain the ground-state of the three-spin cluster, 1$^+$, times the
ground-state for the rest of the (open) chain with two sites removed
resulting in the $1^-$ ground-state.
Eventually, if $J_1$ is kept fixed, the effective $S=1$ impurity is screened
for long enough chains,
producing the $0^+$ ground-state, but this process proceeds
logarithmically slowly. Finite-size scaling analysis, although not very
reliable due to the marginal operators, seems to indicate that $J_2$ is
relevant, as expected.

Now consider what happens as we decrease $J_1$.  At $J_1=0$ we obtain,
approximately the open$^+\times (S=1)$ spectrum, shown in
Table~\ref{spectrum}, as expected.  Note also, the minimum in the $1^+ - 0^-$
gap which occurs for $J_1 \approx .7$.  With increasing $L$ we expect a
crossing of these two levels to occur.  Asymptotically the $0^-$ state
should lie below the $1^+$ state for all $J_1$ such that $0<J_1<1$,
since this region should be in the open$^-$ phase.

We now consider  the effect of varying $J_2$ away from $0$ with $J_1$
held fixed at its integrable value, $J_1=1$; [see Fig.~\ref{fig:numJ2}].
The open$^-$ spectrum of Table~\ref{spectrum} is obtained for large positive
$J_2$, as expected.  Although we expect to be in the open$^+\times(S=1)$
phase for sufficiently large $J_2<0$, this is not obvious from
Fig.~\ref{fig:numJ2}.  The problem is that, for $J_2\to -\infty$ for fixed
$L$ the three-spin complex has an $S=2$ ground-state giving a $2^-$
ground-state including the decoupled open chain.  The screening of this
$S=2$ effective impurity down to $S=1$ is logarithmically slow. Indeed,
the open$^+\times (S=1)$ spectrum is best approximated in this region for
$J_2 \approx -.3$.  This may correspond to crossing the open$^- -
\hbox{open}^+\times (S=1)$ phase boundary, in Fig.~\ref{fig:phase}
where the marginal boundary coupling vanishes.

 According to the ``$g$-theorem'' the
``ground-state degeneracy'', $g$, decreases under renormalization from
less stable to more stable fixed points.\cite{AL1}  The value of $g$ for a
periodic chain\cite{imp1} is $1$ and for an open chain is $1/\sqrt{2}$.
These values must be multiplied by the degeneracy of the decoupled
impurity at each critical point. Thus, at $P_2$, $g=4/\sqrt{2}$; at
$P_1$, $g=2$, at the origin, $g=3/\sqrt{2}$ and at the open$^+$ and
open$^-$ critical points at $\infty$, $g=1/\sqrt{2}$.  We see that, in
all cases, the $g$-theorem is obeyed. A pictorial summary of the five
fixed points and the corresponding values of g are given in
Fig.~\ref{fig:flow}.

\section{Conclusions} The apparent contradiction between the integrable
impurity
model and our RG analysis is explained by the fact that this model
renormalizes to an {\it unstable} critical point corresponding to a
periodic chain with a decoupled $S=1/2$ impurity.  This has been shown
rather convincingly for the integrable
impurity model itself from finite-size
analysis of chains of length $L\leq 5,000$.  A general phase diagram has
been proposed and supported by finite-size analysis of general
models on chains with $L \leq 20$.

\section{Acknowledgements}
We would like to thank Henrik Johannesson for calling our attention to
the work on the integrable
impurity model.  This research was supported in part
by NSERC of Canada.

\begin{table}
\caption{The low temperature spectra showing {\em only} the four
states  $0^+, 0^-, 1^+, 1^-$ for the various fixed points. $L$ is
divisible by 4, $x=L(E_n-E_0)/2\pi v.$}
\label{spectrum} \begin{tabular}{|c|c|c|c|c|}
$x$
&Integrable Model: & Open Chain & Open Chain &  Open  Chain \\
&Periodic Chain& of $L-2$ sites & of $L$ or $L-4$ sites &  of $L$ sites with \\
&of $L+1$ sites with& with singlet & with singlet &  decoupled spin-1 \\
&decoupled spin-1/2 impurity& & & impurity\\
&$P_1$&open$^-$&open$^+$&open$^+\times(S=1)$\\
\tableline \tableline
  7/4 & & $0^+$  & $0^-$ &  \\  \tableline 5/4 & & $1^-$ & $1^+$ & $0^+$ \\
\tableline 3/4 & & $1^+$ & $1^-$ & $0^-, 1^-$ \\ \tableline 1/4 & $0^+, 0^-,
1^+, 1^-$ & $0^-$ & $0^+$ & $1^+$

\end{tabular} \end{table}

\begin{table}
\caption{Spectrum of the integrable impurity model for {\em even}
chains, corresponding to an odd number, $L$,
of sites with S=1/2 plus one spin S=1. The levels shown are the ground-
state, $E_{1/2}$, which has S=1/2, the first excited state, $E_{3/2}$,
which has S=3/2, and the second excited state, $E^*_{1/2}$, which has
S=1/2. }\label{tab:even} \begin{tabular}{rddd} \multicolumn{1}{c}{L}
&\multicolumn{1}{c}{$-E_{1/2}$} &\multicolumn{1}{c}{$-E_{3/2}$}
&\multicolumn{1}{c}{$-E^*_{1/2}$}\\
\tableline
3&        2.666666666667 &          2.000000000000 & 0.833333333333\\
5&        4.000000000000 &          3.535183758488 & 2.788431731592\\
7&        5.360920843433 &          5.000000000000 & 4.473627449978\\
9&        6.732050807569 &          6.435509583680 & 6.035934151803\\
11&       8.108179626193 &          7.855772506636 & 7.536851526445\\
13&       9.487154267776 &          9.267023613165 & 9.003326289075\\
15&      10.867912224042 &         10.672479361373 & 10.448698007398\\
17&      12.249865197376 &         12.073976382835 & 11.880255773559\\
19&      13.632659805114 &         13.472643412759 & 13.302298857367\\
21&      15.016070414566 &         14.869214052794 & 14.717519700013\\
99&      69.058063860390 &         69.021685474014 & 68.996582070614\\
199&    138.369221288792 &        138.350334224364 & 138.339077391494\\
399&    276.996826009204 &        276.987093406252 & 276.981995627127\\
599&    415.625640252833 &        415.619054638227 & 415.615834877143\\
799&    554.254762484295 &        554.249776039106 & 554.247448415855\\
999&    692.884009113116 &        692.879992507487 & 692.8781812698\\
2047&  1419.301863049853 &       1419.299863460628 & -\\
4999&  3465.472116743763 &       3465.471281523977 & -\\
\end{tabular}
\end{table}

\begin{table} \caption{Spectrum of the integrable impurity model for
{\em odd} chains, corresponding to an even number, $L$,
of sites with S=1/2 plus one spin S=1. The levels shown are the ground-
state, $E_{1}$, which has S=1,
and the second excited state, $E_{2}$, which has S=2.
}\label{tab:odd} \begin{tabular}{rdd}
\multicolumn{1}{c}{L}
&\multicolumn{1}{c}{$-E_{1}$}
&\multicolumn{1}{c}{$-E_{2}$}\\
\tableline
2    &     1.333333333333 &  0.000000000000 \\
4    &     2.951367322083 &  1.767591879244 \\
6    &     4.413722666901 &  3.489392593998 \\
8    &     5.839581489564 &  5.090273141761 \\
10   &     7.250399759403 &  6.621632325853 \\
12   &     8.653468913494 &  8.111911768408 \\
14   &    10.051996142025 &  9.576268507776 \\
16   &    11.447623604652 &  11.023300473744 \\
18   &    12.841282703621 &  12.458202998467 \\
20   &    14.233542424308 &  13.884288605831 \\
100  &    69.731013913337 &  69.651755717692 \\
200  &   139.052006805828 & 139.011131675689 \\
400  &   277.684686429578 & 277.663763091977 \\
600  &   416.315227987524 & 416.301121655718 \\
800  &   554.945223280210 & 554.934568453894 \\
1000 &   693.574997489704 & 693.566430520955 \\
2046 &  1418.607647701090 &1418.603401388401 \\
\end{tabular}
\end{table}

\begin{table}
\caption{
The quantum numbers $S_L$, $S_R$, $S_{\rm chain}$, $\vec S_L \cdot \vec
S_R$,
$\vec S_{\rm chain} \cdot \vec S_{\rm eff}$, and $x$ for the five
levels, describing the integrable impurity model.
}\label{tab:states}
\begin{tabular}{ccccccc}
\multicolumn{1}{c}{$S_T^P$}
&\multicolumn{1}{c}{$S_L$}
&\multicolumn{1}{c}{$S_R$}
&\multicolumn{1}{c}{$S_{\rm chain}$}
&\multicolumn{1}{c}{$\vec S_L \cdot \vec S_R$}
&\multicolumn{1}{c}{$\vec S_{\rm chain} \cdot \vec S_{\rm eff}$}
&\multicolumn{1}{c}{$x$}\\
\tableline
L+1 Odd & & & & & & \\
$2^+,2^-$ & 1 & ${1}/{2}$ & ${3}/{2}$ & ${1}/{2}$ & ${3}/{4}$ &
${5}/{4}$ \\
\tableline
$0^+,0^-$ & ${1}/{2}$& 0 & ${1}/{2}$ & 0 & -${3}/{4}$ & ${1}/{4}$\\
\tableline
$1^+,1^-$ & ${1}/{2}$& 0 & ${1}/{2}$ & 0 & ${1}/{4}$ & ${1}/{4}$\\
\tableline
L+1 Even & & & & & & \\
${1}/{2}^{*+}$ & ${1}/{2}$ & ${1}/{2}$ & 0 & -${3}/{4}$ & 0
& ${1}/{2}$\\
\tableline
${1}/{2}^{*-}$ & ${1}/{2}$ & ${1}/{2}$ & 1 &  ${1}/{4}$ & -1
& ${1}/{2}$\\
\tableline
${3}/{2}^-$ & ${1}/{2}$ & ${1}/{2}$ & 1 & ${1}/{4}$ &
${1}/{2}$ & ${1}/{2}$\\
\tableline
${1}/{2}^+$ & 0 & 0 & 0 & 0 & 0 & 0 \\
\end{tabular}
\end{table}

\newpage
\begin{figure}
\caption{The $S=1$ impurity effectively ``splits in
half'', donating an extra $S=1/2$ impurity to the
chain.}\label{fig:split} \end{figure}

%\begin{figure}
%\caption{ The scaled energy gap,  $(E_0-E_1)2L/\pi^2$, for odd chain
%lengths as a function of total chain length, $L+1$.} \label{fig:gap}
%\end{figure}

\begin{figure}
\caption{ The coupling constant, $\lambda_i$, for the five levels,
$E_{1/2}, E_{3/2}$, $E_{1/2}^{*-}$, $E_1$ and $E_2$, as a function of
the effective chain length, $l=L+1$.} \label{fig:couplings} \end{figure}

\begin{figure} \caption{ The average
of the five coupling constants, $\lambda_{\rm av}$,
$\lambda=(\lambda_{1/2}+\lambda_{3/2}+
\lambda_{1/2}^{*-}+\lambda_1+\lambda_2)/5$, as a function of the
effective length, $l=L+1$,
compared to the one-loop renormalization group
prediction, $\lambda_{\rm rg}=\lambda_{\rm av}(l_0)/(1+\lambda_{\rm
av}(l_0)\ln(l/l_0))$. In this plot we have used $l_0=1000$.}
\label{fig:beta} \end{figure}

\begin{figure}
\caption{ The coupling $g_{1/2}^{+*}$
and the average of the
coupling, $g_{\rm av}$, from Ref.~\protect\onlinecite{jphysa}, as a
function of the effective chain length, $l=L+1$,
compared to the one-loop renormalization group
prediction,
$g_{\rm rg}=g_{\rm av}(l_0)/(1+ 4\pi
g_{\rm av}(l_0)\ln(l/l_0)/\protect\sqrt{3})$. In this plot we have used
$l_0=1000$.}
\label{fig:betag} \end{figure}

\begin{figure}\caption{Schematic drawing of the ground-state
wave-functions for the three-spin system depending on $J_i$. ``0"
denotes the $S^z=0$ state of the $S=1$ impurity.}
\label{fig:3spin}\end{figure}

\begin{figure} \caption{The conjectured
phase diagram for the general Hamiltonian, Eq.~(\protect\ref{Hamgen}),
in the parameter
space $J_1, \ J_2$. The couplings $\lambda,\lambda'$ are defined in the
vicinity of
the multi-critical point $P_1$ in Eq.~(\protect\ref{eq:4.2}).
The couplings $e,e_i$ are defined in the vicinity of the multi-critical
fixed point, $P_2$, in Eq.~(\protect\ref{eq:4.3}).}
\label{fig:phase} \end{figure}

%\begin{figure} \caption{Phase diagram for a periodic chain coupled to an
%$S=1/2$ impurity.}\label{fig:phaseP1}\end{figure}

\begin{figure}\caption{
$\vec S_{\rm eff}$ couples with strength $J_2$ to the
two nearest neighbour spins, $\vec \sigma_1$ and $\vec \sigma_L$.
Antiferromagnetic $J_2$ leads to  screening of $\vec S_{\rm eff}$ by three
chain spins and hence to the open$^-$ phase.}\label{fig:nns}\end{figure}

\begin{figure}
\caption{Scaled energy gaps, $L(E-E_{1^+})/2\pi v$, as a function of $J_2$
($J_1 = 1$, $L= 20$).}
\label{fig:numJ2} \end{figure}

\begin{figure}\caption{The unstable fixed point $P_2$ occurring in the
phase diagram, Fig.~\protect\ref{fig:phase}.}\label{fig:P2}\end{figure}

\begin{figure}
\caption{Scaled energy gaps, $L(E-E_{1^+})/2\pi v$,
as a function of $J_1$ ($J_2 = 0, \ L=8,20$).}
\label{fig:numJ1}
\end{figure}
\begin{figure}
\caption{The five fixed points and the RG flows occurring in
Fig.~\protect\ref{fig:phase}. ``0"
denotes the $S^z=0$ state of the $S=1$ impurity.}
\label{fig:flow} \end{figure}

\end{document}